\documentclass[aps,twocolumn,showpacs,preprintnumbers]{revtex4}

\usepackage{graphicx}
\usepackage{dcolumn}
\usepackage{bm}
\usepackage{hyperref}

\newcommand{\beq}{\begin{equation}}
\newcommand{\eeq}{\end{equation}}
\newcommand{\beqa}{\begin{eqnarray}}
\newcommand{\eeqa}{\end{eqnarray}}
\newcommand{\beqar}{\begin{eqnarray*}}
\newcommand{\eeqar}{\end{eqnarray*}}
\newcommand{\bra}[1]{\mbox{$\langle{#1}|$}}
\newcommand{\ket}[1]{\mbox{$|{#1}\rangle$}}

\def\Tr{{\rm Tr}}

\newcounter{saveeqn}
\newcommand{\alpheqn}{\setcounter{saveeqn}{\value{equation}}%
\stepcounter{saveeqn}\setcounter{equation}{0}%
\renewcommand{\theequation}{\mbox{\arabic{saveeqn}\alph{equation}}}}
\newcommand{\reseteqn}{\setcounter{equation}{\value{saveeqn}}%
\renewcommand{\theequation}{\arabic{equation}}}
\def\beql{\alpheqn \beqa}
\def\eeql{\eeqa \reseteqn}

\begin{document}

\title{Partially separable criterion and classification of states\\ in  multipartite systems with an arbitrary finite number of qubits}
\author{
{\large (\em Revised Version)}\\[10pt]
An Min WANG$^{1,2}$}

\altaffiliation{It is founded by the National Fundamental Research Program of China with No. 2001CB309310, partially supported by the National Natural Science Foundation of China under Grant No. 60173047  and the Natural Science Foundation of Anhui Province}

\affiliation{$^{1}$ Laboratory of Quantum Communication and Quantum Computing
and Institute for Theoretical Physics}
\affiliation{$^{2}$Department of Modern Physics,
University of Science and Technology of China \\
P.O. Box 4, Hefei 230027, People's Republic of China
}

\begin{abstract}
After introducing the partially separable concept, we proved the equivalence between the partial separability of a given $m$-partite subsystem with $m$ qubits and the purity of states of this $m$-partite subsystem for a pure state in multipartite systems with arbitrary finite $n(>m)$ qubits. Furthermore, we give out the operational realizations (corollaries) of our theorem, which are the sufficient and necessary criterions of partial separability of states and can be used to classification of states.  Our results are helpful to understand and describe quantum entanglement in multipartite systems. 
\end{abstract}

\pacs{03.65.Ud  03.67.-a }

\maketitle

In order to understand the possibly different ways that a multipartite system can be entangled, one would like to study the classification of states. Up to with four qubits, some various classifications of states have been proposed \cite{Dur1,Verstraete,Luque,Miyake}. However, in the general cases more than four qubits, this problem still keeps unclear and unexpected complicated, and even is thought to be opened \cite{Luque} although the ref.\cite{Miyake}, which was found after we finished our paper, indeed provided a helpful method. 

In our point of view, since the problem of partial separability of even pure states in multipartite systems is not solved clearly, the task of classification of states be, in general, thought to be very complicated. Therefore, we hope to find the operational criterions to classify states by study of the partially separable property of the states. In fact, in our previous paper, we had ever considered it by a partially separable criterion of pure states\cite{My1}. From our physical intuition, we feel that this problem should be able to be fixed in a simpler way and from a physical condition. In this paper, our aim is just to do these. We prove the equivalence between the partial separability of a given $m$-partite subsystem with $m$ qubits and the purity of states of this $m$-partite subsystems for a pure state in multipartite systems with arbitrary finite $n (>m)$ qubits. Furthermore, we give out the operational realizations (corollaries) of our theorem, which are the sufficient and necessary criterions of partial separability of states and can be used to classification of states. 
   
Let us start with the definition of separability of states. 

In the density matrix formalism, a $n$-partite quantum state $\rho_{A_1\!A_2\!\cdots\! A_n}$ is called (fully) separable iff it can be written as a convex combination of product states, {\it i.e.}\cite{Peres,Horodecki1}
\beq
\label{sdef}
\rho_{A_1\!A_2\!\cdots\! A_n}=\sum_m p_m \rho_{A_1}^{(m)}\otimes\rho_{A_2}^{(m)}\otimes\cdots\otimes\rho_{A_n}^{(m)}
\eeq
where any $\rho_{A_i}^{(m)}\  (i=1,2,\cdots,n)$ is a $r_i$-dimensional density matrix. Otherwise, $\rho_{A_1\!A_2\!\cdots\! A_n}$ is entangled or non (fully) separable. In special, if a pure state $\rho_{A_1\!A_2\!\cdots\! A_n}^{\rm P}$ is fully separable, thus it must be ``purely" separable, that is
\beq
\label{scps}
\rho_{A_1\!A_2\!\cdots\! A_n}^{\rm P}=\ket{\psi_{A_1\!A_2\!\cdots\! A_n}}\bra{\psi_{A_1\!A_2\!\cdots\! A_n}}=\prod_{i=1,\otimes}^n\rho_{A_i}^{\rm P}
\eeq 

Now, we would like to introduce a partially separable concept of states. It only exists in more than tripartite systems. As an example, considering a pure state in a tripartite system with three qubits, that $A_3$-part (the third qubit) is separable with $A_1A_2$ parts (a 2-partite subsystem with the first and second qubits) and that $A_1$ part (the first qubit) is separable with $A_2A_3$ parts (a 2-partite subsystem with the second and third qubits) imply simply
\beql
\rho_{A_1\!A_2\!A_3}^{\rm P}=\rho_{A_1\!A_2}^{\rm P}\otimes\rho_{A_3}^{\rm P}\\
\rho_{A_1\!A_2\!A_3}^{\rm P}=\rho_{A_1}^{\rm P}\otimes\rho_{A_2\!A_3}^{\rm P}
\eeql
However, if $A_2$ part (the second qubit) is separable with $A_1A_3$ parts (a 2-partite subsystem with the first and third qubits), the case seems to be not so simple. In order to understand it, we first define an exchanging operator ($4\times 4$ dimensional) for the product of near two 2-dimensional matrices 
\beq
\label{change2q}
S=\left(\begin{array}{cccc}
1&0&0&0\\
0&0&1&0\\
0&1&0&0\\
0&0&0&1
\end{array}\right)
\eeq
It is easy to see that
\beq
S^{-1}=S=S^\dagger
\eeq
That is that it is unitary and Hermit. For any two $2\times 2$ dimensional matrices $M_1$ and $M_2$, it is easy to verify that the action of exchanging operator is
\beq
S(M_1\otimes M_2)S^{-1}=M_2\otimes M_1
\eeq
Obviously, for the direct product of $n$ $2\times 2$ dimensional matrices $M_1,M_2,\cdots,M_n$, we always can use the product of a serial of exchanging operators for near two matrices to rearrange an arbitrary matrix to any position. So it can be called as a rearranged transformation. For example, the $i$th-position matrix can be changed to the end by a following rearranged transformation
\beq
\label{rea}
{\cal{S}}^{(n)}_{i,n}=S^{(n)}_{n-1,n}S^{(n)}_{n-2,n-1}\cdots S^{(n)}_{i,i+1}
\eeq
where in $2^n$ dimensional space
\beq
S^{(n)}_{j,j+1}=I_1\otimes I_2\otimes\cdots\otimes I_{j-1}\otimes S\otimes I_{j+2}\otimes\cdots\otimes I_{n}
\eeq
Here, $I_k$ means a $2\times 2$ unit matrix at $k$ part. Under this rearranged transformation, we have 
\beqa
& &{\cal{S}}^{(n)}_{i,n}M_1\!\otimes\! M_2\otimes\!\cdots\!\otimes M_n{\cal{S}}_{i,n}^{{(n)}-1}\nonumber\\
& &\quad =M_1\!\otimes\! M_2\!\otimes\!\cdots\!\otimes\! M_{i-1}\!\otimes M_{i+1}\!\otimes\!\cdots\! M_n\!\otimes\! M_i
\eeqa

Thus, that $A_2$ part is separable with $A_1A_3$ parts means
\beq
\label{rea3q}
{\cal{S}}^{(3)}_{2,3}\rho_{A_1\!A_2\!A_3}^{\rm P}{\cal{S}}_{2,3}^{{(3)}-1}=\rho_{A_1A_3}^{\rm P}\otimes\rho_{A_2}^{\rm P}
\eeq
where $\rho_{A_1\!A_3}^{\rm P}$ is a reduced density matrix by partially tracing off $A_2$ part, and $\rho_{A_2}^{\rm P}$ is a reduced matrix by partially tracing off $A_1\!A_3$ parts. It must be emphasized that we have used the fact that the rearranged transformation does not change quantum state purity and so we have our lemma as the following:

{\bf Lemma}\ The reduced density matrices including only rearranged parts or no  rearranged parts of a state are invariant under a rearranged transformation. 

{\bf Proof} Just well known, the density matrix  $\rho_{A_1\!A_2\!\cdots\! A_n}$ in a $n$-partite system with $n$ qubits can be expanded as
\beqa
\label{eform}
\rho_{A_1\!A_2\!\cdots\! A_n}&=&\frac{1}{2^n}\sum_{\mu_1=0}^{3}\sum_{\mu_2=0}^{3}\cdots\sum_{\mu_n=0}^{3} \nonumber\\
& &a_{\mu_1\mu_2\cdots \mu_n} \sigma_{A_1}^{\mu_1}\otimes\sigma_{A_2}^{\mu_2}\otimes\cdots\otimes\sigma_{A_n}^{\mu_n}
\eeqa
where $a_{\mu_1\mu_2\cdots \mu_n}$ is a $n$-rank real tensor, $\sigma_{A_i}^0$ is 2-dimensional identity matrix and $\sigma^k_{A_i} (k=1,2,3)$ are Pauli matrices.  Thus, under a rearranged transformation ${\cal{S}}^{(n)}_{i,n}$ for one part $A_i$, we have
\beqa
\label{rho1pra}
& &\rho_{A_1\!A_2\!\cdots\!A_{i-1}\!A_{i+1}\!\cdots\!A_n\!A_i}^\prime ={\cal{S}}^{(n)}_{i,n}\rho_{A_1\!A_2\!\cdots\!A_n}{\cal{S}}_{i,n}^{{(n)}-1}\nonumber\\
& &=\frac{1}{2^n}\sum_{\mu_1=0}^{3}\sum_{\mu_2=0}^{3}\cdots\sum_{\mu_n=0}^{3} a_{\mu_1\mu_2\cdots \mu_n} \sigma_{A_1}^{\mu_1}\otimes\sigma_{A_2}^{\mu_2}\nonumber\\
& &\quad \otimes\cdots \sigma^{\mu_{i-1}}_{A_{i-1}}\otimes\sigma^{\mu_{i+1}}_{A_{i+1}}\cdots\otimes\sigma_{A_n}^{\mu_n}\otimes\sigma^{\mu_i}_{A_i}
\eeqa
It is easy to see that
\beq
\Tr_{A_i}(\rho_{A_1\!A_2\!\cdots\!A_{i-1}\!A_{i+1}\!\cdots\!A_n\!A_i}^\prime)=\Tr_{A_i}(\rho_{A_1\!A_2\!\cdots\! A_n})
\eeq
\beqa
& &\Tr_{\prod_{j=1,j\neq i}^n A_j}(\rho_{A_1\!A_2\!\cdots\!A_{i-1}\!A_{i+1}\!\cdots\!A_n\!A_i}^\prime) \nonumber\\
& &\quad =\Tr_{\prod_{j=1,j\neq i}^n A_j}(\rho_{A_1\!A_2\!\cdots\! A_n})
\eeqa
that is, 
\beql
\rho_{A_1\!A_2\!\cdots\!A_{i-1}\!A_{i+1}\!\cdots\!A_n}^\prime&=&\rho_{A_1\!A_2\!\cdots\!A_{i-1}\!A_{i+1}\!\cdots\!A_n}\\
\rho_{A_i}^\prime&=&\rho_{A_i}
\eeql
Further, from the rearrangement of two parts $A_i$ and $A_j$ $(i<j)$ , 
\beqa
& &{\cal{S}}^{(n)}_{i,n-1}{\cal{S}}^{(n)}_{j,n}\rho_{A_1\!A_2\!\cdots\! A_n}{\cal{S}}_{j,n}^{{(n)}-1}{\cal{S}}_{i,n-1}^{{(n)}-1}\nonumber\\
& &=\frac{1}{2^n}\sum_{\mu_1=0}^{3}\sum_{\mu_2=0}^{3}\cdots\sum_{\mu_n=0}^{3} a_{\mu_1\mu_2\cdots \mu_n} \nonumber\\
& &\quad  \sigma_{A_1}^{\mu_1}\otimes\sigma_{A_2}^{\mu_2}\otimes\cdots\sigma^{\mu_{i-1}}_{A_{i-1}}\otimes\sigma^{\mu_{i+1}}_{A_{i+1}}\cdots\nonumber\\
& &\quad \sigma^{\mu_{j-1}}_{A_{j-1}}\otimes\sigma^{\mu_{j+1}}_{A_{j+1}}\cdots\otimes\sigma_{A_n}^{\mu_n}\otimes\sigma^{\mu_i}_{A_i}\otimes^{\mu_j}_{A_j}
\eeqa
it follows that
\beql
& &\rho^\prime_{A_1\!A_2\!\cdots\!A_{i-1}\!A_{i+1}\!\cdots\!A_{j-1}\!A_{j+1}\!\cdots\!A_n}\nonumber\\
& & \qquad= \rho_{A_1\!A_2\!\cdots\!A_{i-1}\!A_{i+1}\!\cdots\!A_{j-1}\!A_{j+1}\!\cdots\!A_n}\\
& &\rho_{A_i\!A_j}^\prime =\rho_{A_i\!A_j}
\eeql
Likewise, we can prove our lemma in the cases of rearrangement of $m$-parts. Moreover, it is valid both in the pure and in the mixed states.   

Thus, a pure state in a $n$-partite system with $n$ qubits is defined as one part, for example $A_i$ ($i$ can takes arbitrary one value from 1 to $n$), partially separable if it has the following behavior
\beq
\label{1psps}
{\cal{S}}^{(n)}_{i,n}\rho_{A_1\!A_2\!\cdots\! A_n}^{\rm P}{\cal{S}}_{i,n}^{{(n)}-1}=\rho_{A_1\!A_2\!\cdots\!A_{i-1}A_{i+1}\cdots  A_n}^{\rm P}\otimes\rho_{A_i}^{\rm P}
\eeq 
{\it via.} a rearrangement transformation (\ref{rea}). 

Obviously, {\em if any part ($i$ takes over all values) is one part partially separable, then this pure state is fully separable.}

Likewise, a 2-partite subsystem with $A_iA_j$ $(i<j)$ parts is defined as two part partially separable by 
\beqa
\label{2psps}
& &{\cal{S}}^{(n)}_{i,n-1}{\cal{S}}^{(n)}_{j,n}\rho_{A_1\!A_2\!\cdots\! A_n}^{\rm P}{\cal{S}}_{j,n}^{{(n)}-1}{\cal{S}}_{i,n-1}^{{(n)}-1}\nonumber \\
& &\quad =\rho_{A_1\!A_2\!\cdots\!A_{i-1}A_{i+1}\cdots\!A_{j-1}A_{j+1}\cdots  A_n}^{\rm P}\otimes\rho_{A_i\!A_j}^{\rm P}
\eeqa
It must be emphasized that it will be a suitable classification if in the separated subsystem $A_i$ part and $A_j$ part are entangled, that is, $\rho_{A_iA_j}^{\rm P}$ is an entangled state between $A_i$ part and $A_j$ part. Otherwise both $A_i$ and $A_j$ will be one part partially separable. It implies that $A_i$ and $A_j$ parts ought not to be brought into two part partially separable classes.

Of course, we can introduce $m$-part partially separable concept of states in a similar way. It must be emphasized that {\em it is enough only considering up to $[n/2]$ part partially separable cases for a $n$-partite system, where $[\cdots]$ means to take integral part}. This is because if a pure state in $n$-partite systems is $m$ part partially separable, then it is also $n-m$ part partially separable, at least in form. But, in two cases the separated parts which are referred are different.  

It is easy to see that {\em if no any part(s) (including one, two, $\cdots$ up to $[n/2]$) is partially separable, then this pure state is fully entangled among all of parts.}

The concept of partial separability can be extended to the case of mixed states. If a mixed state can be written as such a pure state decomposition $\sum_a p_a\rho_a$ ($\rho^a$ is pure) that 
\beq
\label{1psms}
{\cal{S}}^{(n)}_{i,n}\rho_{A_1\!A_2\!\cdots\! A_n}{\cal{S}}_{i,n}^{{(n)}-1}=\sum_a p_a\rho_{A_1\!A_2\!\cdots\!A_{i-1}A_{i+1}\cdots  A_n}^{a}\otimes\rho^a_{A_i}
\eeq 
it is just one part partially separable. Obviously, if for every pure state $\rho^a$, any part is one part partially separable, this mixed state is just fully separable. Therefore, {\em the fully separability is an extreme case of partial separability}. Through study of partial separability, one can obtain not only the knowledge of fully separability, but also the tools of classification of states. 

For a mixed state, the two part partially separable case is with form
\beqa
\label{2psms}
\!\!\!\!& &{\cal{S}}^{(n)}_{i,n-1}{\cal{S}}^{(n)}_{j,n}\rho_{A_1\!A_2\!\cdots\! A_n}{\cal{S}}_{j,n}^{{(n)}-1}{\cal{S}}_{i,n-1}^{{(n)}-1}\nonumber\\
& &=\sum_a p_a\rho_{A_1\!A_2\!\cdots\!A_{i-1}A_{i+1}\!\cdots\! A_{j-1}A_{j+1}\!\cdots\! A_n}^{a}\otimes\rho_{A_i\!A_j}^a
\eeqa 
Similarly, there is the concept of partial separability of a given number of parts for mixed states.

In addition, we have the ``purely" partially separable concept of mixed states, that is, a mixed state is one part and two part purely partially separable if  one has
\beq
\label{1ppsms}
{\cal{S}}^{(n)}_{i,n}\rho_{A_1\!A_2\!\cdots\! A_n}{\cal{S}}_{i,n}^{{(n)}-1}=\left(\sum_a p_a\rho_{A_1\!A_2\!\cdots\!A_{i-1}A_{i+1}\cdots  A_n}^{a}\right)\otimes\rho_{A_i}
\eeq
and
\beqa
\label{2ppsms}
\!\!\!\!\!\!\! & &{\cal{S}}^{(n)}_{i,n-1}{\cal{S}}^{(n)}_{j,n}\rho_{A_1\!A_2\!\cdots\! A_n}{\cal{S}}_{j,n}^{{(n)}-1}{\cal{S}}_{i,n-1}^{{(n)}-1}\nonumber\\
\!\!\!\!\!\!\!& &=\left(\sum_a p_a\rho_{A_1\!A_2\!\cdots\!A_{i-1}A_{i+1}\!\cdots\! A_{j-1}A_{j+1}\!\cdots\! A_n}^{a}\right)\!\otimes\!\rho_{A_i\!A_j}
\eeqa
respectively. Similarly, there is the concept of purely partial separability of a given number of parts for mixed states.

Now, we can express our central theorem:

{\bf Theorem} For a pure state in $n$-partite systems with $n$ qubits (every part has a qubit), the partial separability of a given $m$-partite subsystem with $m (<n)$ qubits is equivalent to the purity of states of this $m$-partite subsystem.

{\bf Proof} First, let us prove, for a pure state in $n$-partite systems with $n$ qubits (every part has a qubit), if the state of a given $m$-partite subsystem with $m(< n)$ qubits is separable with the other $(n-m)$-partite subsystem with $n-m$ qubits, it must be a pure state. This proof is direct and simple after introducing the partially separable concept and using our lemma. From the concept of partial separability of a given $m$-partite subsystem and based our lemma, we always can write out the expression of partial separability in a rearranged transformation ${\cal{S}}$, that is  
\beq
{\cal{S}}\rho_{A_1\!A_2\!\cdots\! A_n}^{\rm P}{\cal{S}}=\rho_{A_{i_1}A_{i_2}\cdots A_{i_{n-m}}}\otimes\rho_{A_{j_1}A_{j_2}\cdots A_{j_{m}}} 
\eeq
where all of $A_{j_1},A_{j_2},\cdots, A_{j_{m}}$ parts form a given separated subsystem. Obviously, from the definition of the rearranged transformation, we know that it does not change the purity of state. Thus, it implies that $\rho_{A_{j_1}A_{j_2}\cdots A_{j_{m}}}$ must be pure. In other words, the purity of states of a given $m$-partite subsystem is a necessary condition of this $m$-partite subsystem to be partially separable with the other $(n-m)$-partite subsystem. 

Actually, for our theorem, it is the key matter to prove, for a pure state in $n$-partite systems with $n$ qubits, if the state of a given $m$-partite subsystem is pure, it will be partially separable with the other $(n-m)$-partite subsystem. In other words, the purity of states of the given $m$-partite subsystem is a sufficient condition of this $m$-partite subsystem to be  partially separable with the other $(n-m)$-partite subsystem. 

In order to prove this sufficiency, we need to use the concept of ``coherent vector" which is defined, for a $r$-dimensional density matrix $\rho(r)$, by the relation
\beq
\label{dcv}
\rho(r)=\frac{1}{r}I_{r\times r}+\frac{1}{2}\sum_{s=1}^{r^2-1}{\xi}^{k}{\lambda}_{s}
\eeq
where $I_{r\times r}$ is $r$-dimensional identity matrix and ${\lambda}_{s}\ (s=1,2,\cdots, r^2-1)$ are generators of $SU(r)$. All of $\{\xi^s,\;s=1,2,\cdots, r^2-1\}$, as components, form a coherent vector. In special, in the simplest 2-dimensional case (one qubit), for example $A_i$ part, the coherent vector is just the polarized vector $\bm{\xi}_{A_i}$ of the reduced density matrix $\rho_{A_i}=\Tr_{\prod_{j=1,j\neq i}^n A_j}(\rho_{A_1\!A_2\cdots A_n})=\frac{1}{2}(\sigma_0+\bm{\xi}_{A_i}\cdot\bm{\sigma})$ (tracing off $n-1$ parts). That is
\beq
\label{pvemp}
\bm{\xi}_{A_i}=\Tr(\rho_{A_1\!A_2\!\cdots\! A_n}I_1\otimes\cdots\otimes I_{i-1}\otimes\bm{\sigma}\otimes I_{i+1}\otimes\cdots\otimes I_n)
\eeq
In four dimensional case (two qubits), the components of the coherent vector of reduced density matrix $\rho_{A_i\!A_j}$ have 15. 
Here, we prefer to choose the tensor product of Pauli matrices as the basis $\lambda_s$ for even dimensional cases, because we deal with the density matrix of multipartite systems made up of qubits. For example, for a system with $m$ qubits, $\lambda_s$ can be taken as  
\beq
\label{basiscv}
\frac{1}{\sqrt{2^{m-1}}}\sigma_{\mu_1}\otimes\sigma_{\mu_2}\otimes\cdots\otimes\sigma_{\mu_m}
\eeq
except for all of $\mu_i\;(i=1,2,\cdots,m)$ are zero at the same time. Here, every $\mu_i$ takes $0,1,2,3$. Of course, $\lambda_s$'s number is $2^{2m}-1$.  
By use of
\beq
\Tr\lambda_s=0,\quad \Tr(\lambda_s\lambda_{t})=2\delta_{st}
\eeq
where $s,t=1,2,\cdots,2^{2m}-1$, we can write down the relation among the components of coherent vector and the generalized Stokes' parameters \cite{Luque}. In a system with 2 qubits as above $\rho_{A_i\!A_j}$, the coherent vector is consist of 
$\{\xi^{k0}_{A_iA_j}$, $\xi^{0l}_{A_iA_j}$, $\xi^{kl}_{A_iA_j}$, $k,l=1,2,3\}$. 
These components can be calculated by
\beqa
\label{pve2q}
& &{\xi}_{A_iA_j}^{\mu_i\mu_j}=\frac{1}{\sqrt{2}}\Tr(\rho_{A_1\!A_2\!\cdots\! A_n}I_1\otimes\cdots
\otimes I_{i-1}\otimes{\sigma}^{\mu_i}\otimes I_{i+1}\nonumber\\
& &\quad \otimes\cdots\otimes I_{j-1}\otimes{\sigma}^{\mu_j}\otimes I_{j+1}\otimes\cdots\otimes I_n)\\
& &\quad =\frac{1}{\sqrt{2}}a_{0\cdots 0 \mu_i 0\cdots 0 \mu_j 0\cdots 0}
\eeqa
where $\mu_i,\mu_j=0,1,2,3$, but $\mu_i$ and $\mu_j$ are not equal to zero at the same time, and $a$'s subscripts are not equal to zero only at the $i$-th and the $j$-th positions. 

Particularly, we have an important conclusion that for a pure state with $m$ qubits, since 
\beq
\label{pscondition}
\Tr\rho^{\rm P 2}_{A_{j_1}A_{j_2}\cdots A_{j_m}}=\frac{1}{2^m}+\frac{1}{2}\bm{\xi}_{A_{j_1}A_{j_2}\cdots A_{j_m}}^2=1
\eeq
The square of norm of the coherent vector is then found. In fact, {\em the above equation (\ref{pscondition}) is a necessary and sufficient condition of purity of state $\rho_{A_{j_1}A_{j_2}\cdots A_{j_m}}$.} 

For the pure states of two qubits, 
\beq
\ket{\psi_{A_1\!A_2}}= a\ket{00}+b\ket{01}+c\ket{10}+d\ket{11}
\eeq
if $\rho_{A_2}$ is pure, it means that the norm its polarized vector $\bm{\xi}_{A_2}$ is 1, again from
\beq
\bm{\xi}_{A_2}^2=1-4 |ad-bc|^2=\bm{\xi}_{A_1}^2
\eeq
it follows
\beq
|ad-bc|=0
\eeq
Actually, it has been well-known as a sufficient condition that $A_2$ part is separable with $A_1$ part, or that $A_1$ part is separable with $A_2$ part \cite{My2}. 

In the case more than two qubits, partially separable situations appear. This problem seems to get a little complicated. 

\begin{widetext}
Consider a pure state in tripartite systems denoted by
\beq
\label{psi3q}
\ket{\psi_{A_1\!A_2\!A_3}}\!\!
=a\ket{000}+b\ket{001}+c\ket{010}+d\ket{011}
+e\ket{100}+f\ket{101}+g\ket{110}+h\ket{111}
\eeq
It is easy to evaluate out the norms of polarized vectors of each part
\beql
\label{nxi3pa}
\bm{\xi}_{A_1}^2\!&=&\!1-4|a f - b e|^2-  4|a g - c e|^2-  4|a h - d e|^2-  4|b g - c f|^2-   4|b h - d f|^2-  4|c h - d g|^2\\
\label{nxi3pb}
\bm{\xi}_{A_2}^2\!&=&\!1-4|a d - b c|^2-  4|a g - c e|^2-  4|a h - c f|^2-  4|b g - d e|^2-  4|b h - d f|^2-   4|e h - f g|^2\\
\label{nxi3pc}
\bm{\xi}_{A_3}^2\!&=&\!1-4|a d - b c|^2-  4|a f - b e|^2-  4|a h - b g|^2-  4|c f - d e|^2-  4|c h - d g|^2-  4|e h - f g|^2
\eeql
It is well known that $\bm{\xi}_{A_i}^2=1,(i=1,2,3)$ is sufficient and necessary condition of $\rho_{A_i}$ purity. Then, from the above equations it follows respectively that 
\beql
\label{ps3pa}
a f = b e, \  a g = c e, \  a h = d e, \  
b g = c f, \  b h = d f, \  c h = d g; &\quad&\mbox{ If $\rho_{A_1}$ is pure or $\bm{\xi}_{A_1}^2=1$}\\
\label{ps3pb}
a d = b c, \  a g = c e, \  a h = c f, \  
b g = d e, \  b h = d f, \  e h = f g; &\quad&\mbox{ If $\rho_{A_2}$ is pure or $\bm{\xi}_{A_2}^2=1$}\\
\label{ps3pc}
a d = b c, \  a f = b e, \  a h = b g, \  
c f = d e, \  c h = d g, \  e h = f g. &\quad&\mbox{ If $\rho_{A_3}$ is pure or $\bm{\xi}_{A_3}^2=1$}
\eeql
To our aim, we express the pure states of three qubits as 255 classes of states based on the number of non zero coefficients, and study them step by step. It is simple when only there is a non-zero coefficient in the states since they is fully separable. When there are only two non-zero coefficients, for example, $a$ and $d$, we have
$\bm{\xi}_{A_1}^2=1$, $\bm{\xi}_{A_2}^2=\bm{\xi}_{A_3}^2=1-4|ad-bc|^2$. It is easy to see that $A_1$-part is separable since $a\ket{000}+d\ket{011}=\ket{0}\otimes(a\ket{00}+d\ket{11})$. Again, let us consider $\ket{\psi}=a\ket{000}+d\ket{011}+e\ket{100}+h\ket{111}$, where $a,d,e,h$ are all not equal to zero. From (\ref{nxi3pa}-\ref{nxi3pc}), we obtain \beql
\bm{\xi}_{A_1}^2&=&1-4|ah-de|^2\\
\bm{\xi}_{A_2}^2&=&1-4|a|^2|d|^2-4|d|^2|e|^2-4|a|^2|h|^2-4|e|^2|h|^2\\
\bm{\xi}_{A_3}^2&=&1-4|a|^2|d|^2-4|d|^2|e|^2-4|a|^2|h|^2-4|e|^2|h|^2
\eeql
Set $\bm{\xi}_{A_1}^2=1$. It implies that $ah=de$, and so 
\beqa
\ket{\psi}&=&a\ket{000}+d\ket{011}+c\ket{100}+h\ket{111}=a\ket{000}+d\ket{011}+\frac{ah}{d}\ket{100}+h\ket{111}\\ \nonumber
&=&\ket{0}\otimes(a\ket{00}+d\ket{11})+\frac{h}{d}\ket{1}\otimes(a\ket{00}+d\ket{11})=\left(\ket{0}+\frac{h}{d}\ket{1}\right)\otimes(a\ket{00}+d\ket{11})
\eeqa
Therefore $\bm{\xi}_{A_1}^2=1$ for above state is a sufficient criterion that $A_1$-part is separable with $A_2\!A_3$. Moreover, since $a,d$ are not zero, BC must be entangled (non separable). Likewise, we can prove our theorem for all classes of state. Our results are listed the following table. It also gives out a classification of pure states in tripartite systems with three qubits. 

\begin{center}
\small
\begin{tabular}{c|c|c|c|c|c}
\multicolumn{6}{c}{\normalsize\bf Classification of the pure states in a tripartite system with three qubits}\\ 
\hline
\multicolumn{3}{c|}{\it Non-zero Coefficients} & \it Fully & \it Partially & \it  \\ \cline{1-3}
\it Number & \multicolumn{2}{|c|}{\it Nonzero Components} & \it Separable &\it Separable &\it  \raisebox{1.23ex}[0pt]{Entangled} \\ \hline
One & 8 & All & Yes & A,B,C & No \\ \hline
& &$(a, b);(a,c);(a,e);(b,d)$;& & &  \\  
& 12 &$(b,f);(c,d);(c,g);(d,h)$; & Yes & A,B,C & No \\  
& & $(e,f);(e,g);(f,h);(g,h)$ &  &  & \\ \cline{2-6}
Two & 4 & $(a,d);(b,c);(e,h);(f,g)$ & No & A-part & BC  \\ \cline{2-6}
& 4 & $(a,f);(b,e);(c,h);(d,g)$ & No & B-part & AC \\ \cline{2-6}
& 4 & $(a,g);(b,h);(c,e);(d,f)$ & No & C-part & AB \\ \cline{2-6}
& 4 & $(a,h);(b,g);(c,f);(d,e)$ & No & No & Yes \\ \hline
& & $(a,b,c);(a,b,d);(a,c,d)$;& & &  \\ 
& 8 & $(b,c,d);(e,f,g);(e,f,h)$;& No & A-part & BC \\ 
& & $(e,g,h);(f,g,h)$ & & & \\ \cline{2-6}
& & $(a,b,e);(a,b,f);(a,e,f)$;& & &  \\ 
& & $(b,e,f);(c,d,g);(c,d,h)$; & & &  \\ 
Three & \raisebox{1.23ex}[0pt]{8} & $(c,g,h);(d,g,h)$ & \raisebox{1.23ex}[0pt]{No} & \raisebox{1.23ex}[0pt]{B-part} & \raisebox{1.23ex}[0pt]{AC} \\ \cline{2-6}
& & $(a,c,e);(a,c,g);(a,e,g)$; & & &  \\ 
& & $(b,d,f);(b,d,h);(b,f,h)$; & & &  \\ 
& \raisebox{1.23ex}[0pt]{8} & $(c,e,g);(d,f,h)$ & \raisebox{1.23ex}[0pt]{No} & \raisebox{1.23ex}[0pt]{C-part} & \raisebox{1.23ex}[0pt]{AB} \\ \cline{2-6}
& 32 & The Others & No & No & Yes \\ \hline
&  & $(a,b,c,d)$ & No or Yes if $ad=bc$ &  & BC or No if $ad=bc$  \\
& \raisebox{1.23ex}[0pt]{2} & $(e,f,g,h)$ & No or Yes if $eh=fg$ & \raisebox{1.23ex}[0pt]{A-part} & BC or No if $eh=fg$  \\ \cline{2-6}
&  & $(a,b,e,f)$ & No or Yes if $af=be$ & & AC or No if $af=be$ \\ 
& \raisebox{1.23ex}[0pt]{2} & $(c,d,g,h)$ & No or Yes if $ch=dg$ & \raisebox{1.23ex}[0pt]{B-part} & AC or No if $ch=dg$ \\ \cline{2-6}
&  & $(a,c,e,g)$ & No or Yes if $ag=ce$ & & AB or No if $ag=ce$\\ 
& \raisebox{1.23ex}[0pt]{2} & $(b,d,f,h)$ & No or Yes if $bh=df$ & \raisebox{1.23ex}[0pt]{C-part} & AB or No if $bh=df$ \\ \cline{2-6}
Four &  & $(a,d,e,h)$ &  & No or A-part if $ah=de$ & Yes or BC if $ah=de$  \\ 
& \raisebox{1.23ex}[0pt]{2} & $(b,c,f,g)$ & \raisebox{1.23ex}[0pt]{No} & No or A-part if $bg=cf$ & Yes or BC if $bg=cf$  \\ \cline{2-6}
&  & $(a,c,f,h)$ &  & No or B-part if $ah=cf$ & Yes or AC if $ah=cf$ \\ 
& \raisebox{1.23ex}[0pt]{2} & $(b,d,e,g)$ & \raisebox{1.23ex}[0pt]{No} & No or B-part if $bg=de$ & Yes or AC if $bg=de$ \\ \cline{2-6}
&  & $(a,b,g,h)$ &  &  No or C-part if $ah=bg$ & Yes or AB if $ah=bg$ \\ 
& \raisebox{1.23ex}[0pt]{2} & $(c,d,e,f)$ & \raisebox{1.23ex}[0pt]{No} &  No or C-part if $cf=de$ & Yes or AB if $cf=de$ \\ \cline{2-6}
& 58 & The Others & No & No & Yes \\ \hline
Five & 56 & All & No & No & Yes \\ \hline
& & $(a,b,c,e,f,g)$ & & No or A-part if $af=be,ag=ce$ & Yes or BC if $af=be,ag=ce$ \\ 
& & $(a,b,d,e,f,h)$ & &No or A-part if $af=be,ah=de$ & Yes or BC if $af=be,ah=de$ \\ 
&\raisebox{1.23ex}[0pt]{4} & $(a,c,d,e,g,h)$ & \raisebox{1.23ex}[0pt]{No} & No or A-part if $ag=ce,ah=de$ & Yes or BC if $ag=ce,ah=de$ \\ 
& & $(b,c,d,f,g,h)$ & & No or A-part if $bg=cf,bh=df$ & Yes or BC if $bg=cf,bh=df$ \\ \cline{2-6}
& & $(a,b,c,d,e,g)$ & & No or B-part if $ad=bc,ag=ce$ & Yes or AC if $ad=bc,ag=ce$\\ 
& & $(a,b,c,d,f,h)$ & & No or B-part if $ad=bc,ah=cf$ & Yes or AC if $ad=bc,ah=cf$\\ 
&\raisebox{1.23ex}[0pt]{4} & $(a,c,e,f,g,h)$ & \raisebox{1.23ex}[0pt]{No} & No or B-part if $ag=ce,ah=cf$ & Yes or AC if $ag=ce,ah=cf$\\ 
\raisebox{3.23ex}[0pt]{Six} & & $(b,d,e,f,g,h)$ & & No or B-part if $bg=de,bh=df$ & Yes or AC if $bg=de,bh=df$\\ \cline{2-6}
& & $(a,b,c,d,e,f)$ & & No or C-part if $ad=bc,af=be$ & Yes or AB if $ad=bc,af=be$\\ 
& & $(a,b,c,d,g,h)$ & & No or C-part if $ad=bc,ah=bg$ & Yes or AB if $ad=bc,ah=bg$\\ 
&\raisebox{1.23ex}[0pt]{4} & $(a,b,e,f,g,h)$ & \raisebox{1.23ex}[0pt]{No} & No or C-part if $af=be,ah=bg$ & Yes or AB if $af=be,ah=bg$\\ 
& & $(c,d,e,f,g,h)$ & & No or C-part if $cf=de,ch=dg$ & Yes or AB if $cf=de,ch=dg$\\ \cline{2-6}
& 16 & The Others & No & No & Yes \\ \hline
Seven & 8 & All & No & No & Yes \\ \hline
& & & No or Yes & No or A-part if eq.(\ref{ps3pa}) is valid & Yes or BC if eq.(\ref{ps3pa}) is valid\\ \cline{5-6}
Eight & 1 & $(a,b,c,d,e,f,g)$ & if eq. (\ref{ps3pa}-c) & No or B-part if eq.(\ref{ps3pb}) is valid & Yes or AC if eq.(\ref{ps3pb}) is valid\\ \cline{5-6}
& & & are all valid & No or C-part if eq.(\ref{ps3pc}) is valid & Yes or AB if eq.(\ref{ps3pc}) is valid\\ \hline
\end{tabular}
\end{center}

\end{widetext}

Now we can use mathematical induction to prove the case of $n$-partite systems. Suppose this one part partially separability criterion is sufficient and necessary for $n-1$ partite systems. In the case of $n$-partite systems, set that the square of norm of polarized vector of $A_i$ part is equal to 1, that is, $\bm{\xi}_{A_i}^2=1$. Then, by tracing off any one part $A_j $ except for a given $A_i$ part $(i\neq j$), a state $\rho_{A_1\!A_2\!\cdots\!A_{j-1}\!A_{j+1}\!\cdots\!A_n}$ in a $n-1$ partite system including $A_i$ is obtained. But, this state is, in general, a mixed state. We can write its pure state decomposition as
\beq
\rho_{A_1\!A_2\!\cdots\!A_{j-1}\!A_{j+1}\!\cdots\!A_n}^\prime=\sum_a p_a\rho_{A_1\!A_2\!\cdots\!A_{j-1}\!A_{j+1}\!\cdots\!A_n}^{\prime\;a}
\eeq
where each $\rho^{\prime\;a}$ is a pure state. Then, again tracing off the other parts except for $A_i$, we have
\beq
\rho_{A_i}=\sum_a p_a \rho^a_{A_i}
\eeq
Because $\bm{\xi}_{A_i}^2=1$ implies that $\rho_{A_i}$ is a pure state, it follows that $\rho^a_{A_i}$ is not dependent on its index $a$, or for all values of $a$, $\rho^a_{A_i}=\rho_{A_i}$. In other words, $(\bm{\xi}_{A_i}^a)^2=\bm{\xi}_{A_i}^2=1$. Based on our precondition that theorem one is valid for $n-1$ partite systems, $A_i$ part is one part partially separable in $\rho_{A_1\!A_2\!\cdots\!A_{j-1}\!A_{j+1}\!\cdots\!A_n}^a$ and so $A_i$ part is one part partially separable with all other parts in $\rho_{A_1\!A_2\!\cdots\!A_{j-1}\!A_{j+1}\!\cdots\!A_n}$. That is
\beqa
\label{n-1sc}
& &{\cal{S}}^{(n-1)}_{i,n-1}\rho_{A_1\!A_2\!\cdots\!A_{j-1}\!A_{j+1}\!\cdots\!A_n}{\cal{S}}_{i,n-1}^{{(n-1)}-1}\nonumber\\
& &\quad =\sum_a p_a \rho_{A_1\!A_2\!\cdots\!A_{i-1}\!A_{i+1}\!\cdots\!A_{j-1}\!A_{j+1}\!\cdots\!A_n}^{\prime\;a}\otimes\rho_{A_i}^a\nonumber\\
& &\quad =\left(\sum_a p_a \rho_{A_1\!A_2\!\cdots\!A_{i-1}\!A_{i+1}\!\cdots\!A_{j-1}\!A_{j+1}\!\cdots\!A_n}^{\prime\;a}\right)\otimes\rho_{A_i}\nonumber\\
& &\quad = \rho_{A_1\!A_2\!\cdots\!A_{i-1}\!A_{i+1}\!\cdots\!A_{j-1}\!A_{j+1}\!\cdots\!A_n}^{\prime}\otimes\rho_{A_i}
\eeqa
Because that $j$ can takes over all values except for the given $i$, we can obtain the conclusion that the $A_i$ part is one part partially separable with arbitrary other parts. In other words, for $n$-partite systems with arbitrary finite $n$ qubits, the purity of states of the given $A_i$ is then a sufficient condition of its partial separability with the other parts. If we can not obtain the forms of $A_i$ part partially separable with the other parts for $n$ parties system when the state of $A_i$ part is pure, it must be conflict with eq.(\ref{n-1sc}) and then our precondition. It is impossible.

It must be emphasized two facts. They are (1) The necessary and sufficient condition that a given part $A_i$ is one part partially separable with the other parts is the state purity of $A_i$ part (our theorem); (2) The necessary and sufficient condition of state purity of one part $A_i$ is $\bm{\xi}_{A_i}^2=1$ (see eq.(\ref{pscondition}) ). Therefore, by combining them together we obtain an operational realization of our theorem when only considering one part partially separable case as the following 

{\bf Corollary One} For a pure state in $n$-partite systems with $n$ qubits (every part has a qubit), one part $A_i$ ($i$ is given) partially separable sufficient and necessary condition is 
\beq
\bm{\xi}_{A_i}^2=1
\eeq

It is worthy of pointing out that we choose the expression of our corollary in order to make our partial separability criterion become operational and be able to use to classification of states. That is, for a pure state $\rho_{A_1\!A_2\!\cdots\! A_n}$, by calculating if $\bm{\xi}_{A_i}^2$ is equal to 1 for a given $A_i$ part or not, one can judge if the given part $A_i$ is partially separable with the other parts or not, and then determine the class of this state.  

As above, it is known that the fully separability of a pure state in multipartite systems means that any part is one part partially separable. So we have the necessary and sufficient criterion of fully separability \cite{My1}: 

{\bf Corollary Two}\  The sufficient and necessary conditions for separability of an arbitrary pure state in $n$-partite systems with $n$ qubits (every part has a qubit) are 
\beq
\label{scforqbmp}
\bm{\xi}_{A_1}^2=\bm{\xi}_{A_2}^2=\cdots=\bm{\xi}_{A_n}^2=1
\eeq
where every $\bm{\xi}_{A_i}$ is defined by eq. (\ref{pvemp}). 

Up two now, we still have not finished our proof because in the case more than three qubits, the two part partially separable and more part partially separable
cases can appear. We have to study the (partial) separability of high dimensional systems generally. We need to continue our proof of sufficiency for these complicated cases. Let us consider them from a pure state in the four partite system, which can be denoted by
\beqa
\label{psi4qb}
\ket{\psi_{A_1\!A_2\!A_3\!A_4}}&=&\!\!\!\sum_{\alpha_1,\alpha_2,\alpha_3,\alpha_4=0}^1\!\!\! x_{\alpha_1\alpha_2\alpha_3\alpha_4}\ket{\alpha_1\alpha_2\alpha_3\alpha_4}\\
&=&\sum_{i=1}^{16} x_i\ket{i}
\eeqa
where $\{\ket{i};i=1,2,\cdots,16\}$=$\{\ket{0000}$, $\ket{0001}$, $\ket{0010}$, $\ket{0011}$, $\ket{0100}$, $\ket{0101}$, $\ket{0110}$, $\ket{0111}$, $\ket{1000}$, $\ket{1001}$, $\ket{1010}$, $\ket{1011}$, $\ket{1100}$, $\ket{1101}$, $\ket{1110}$, $\ket{1111}\}$. 
Obviously, two part partially separable cases have six kinds, that is, $\rho_{A_1\!A_2}$,$\rho_{A_1\!A_3}$,$\rho_{A_1\!A_4}$,$\rho_{A_2\!A_3}$,$\rho_{A_2\!A_4}$,$\rho_{A_3\!A_4}$. In fact, the independent cases only have three kinds, $\rho_{A_1\!A_2},\rho_{A_1\!A_3},\rho_{A_1\!A_4}$ since that two parts $A_1A_2$ partially separable with $A_3A_4$ means two part $A_3A_4$ partially separable with $A_1A_2$ and so on. Because of the purity of $\rho_{A_3\!A_4}$, we have 
\beq
\label{ncv4q2psc}
\bm{\xi}_{A_3\!A_4}^2=3/2
\eeq
Through some computations, we can find that the square of norm of coherent vector of $\rho_{A_3\!A_4}$ is
\beqa
& &\bm{\xi}_{A_3\!A_4}^2=\frac{3}{2}- 4|x_1x_6-x_2x_5|^2-4|x_1x_7-x_3x_5|^2\nonumber\\
& &\quad -4|x_1x_8-x_4x_5|^2-4|x_1x_{10}-x_2x_9|^2\nonumber\\
& &\quad -4|x_1x_{11}-x_3x_9|^2-4|x_1x_{12}-x_4x_9|^2\nonumber\\
& &\quad -4|x_1x_{14}-x_2x_{13}|^2-4|x_1x_{15}-x_3x_{13}|^2\nonumber\\
& &\quad -4|x_1x_{16}-x_4x_{13}|^2-4|x_2x_7-x_3x_6|^2\nonumber\\
& &\quad -4|x_2x_8-x_4x_6|^2-4|x_{2}x_{11}-x_3x_{10}|^2\nonumber\\
& &\quad -4|x_2x_{12}-x_4x_{10}|^2-4|x_2x_{15}-x_3x_{14}|^2\nonumber\\ 
& &\quad -4|x_2x_{16}-x_4x_{14}|^2-4|x_3x_8-x_4x_7|^2\nonumber\\
& &\quad -4|x_3x_{12}-x_4x_{11}|^2-4|x_3x_{16}-x_4x_{15}|^2\nonumber\\
& &\quad -4|x_5x_{10}-x_6x_9|^2-4|x_5x_{11}-x_7x_{9}|^2\nonumber\\
& &\quad -4|x_5x_{12}-x_8x_{9}|^2-4|x_{5}x_{14}-x_6x_{13}|^2\nonumber\\
& &\quad -4|x_5x_{15}-x_7x_{13}|^2-4|x_5x_{16}-x_8 x_{13}|^2\nonumber\\
& &\quad -4|x_6x_{11}-x_7x_{10}|^2-4|x_6x_{12}-x_8x_{10}|^2\nonumber\\
& &\quad -4|x_6x_{15}-x_7x_{14}|^2-4|x_6x_{16}-x_8x_{14}|^2\nonumber\\
& &\quad -4|x_7x_{12}-x_8x_{11}|^2-4|x_7x_{16}-x_8x_{15}|^2\nonumber\\
& &\quad -4|x_9x_{14}-x_{10}x_{13}|^2-4|x_9x_{15}-x_{11}x_{13}|^2\nonumber\\
& &\quad -4|x_9x_{16}-x_{12}x_{13}|^2-4|x_{10}x_{15}-x_{11}x_{14}|^2\nonumber\\
& &\quad -4|x_{10}x_{16}-x_{12}x_{14}|^2-4|x_{11}x_{16}-x_{12}x_{15}|^2
\eeqa
From eq. (\ref{ncv4q2psc}), it follows that there are 36 relations between the pairs of coefficients, that is, all of the squares of norm in the above equation are zero. Then, we can divide the coefficients into four sets: $\bm{v}_1=\{x_1,x_5,x_9,x_{13}\}$; $\bm{v}_2=\{x_2,x_6,x_{10},x_{14}\}$; $\bm{v}_3=\{x_3,x_7,x_{11},x_{15}\}$; $\bm{v}_4=\{x_4,x_8,x_{12},x_{16}\}$, and the component states are also divide into four sets:
\beql
\ket{\bm{\phi}_1}=\{\ket{0000},\ket{0100},\ket{1000},\ket{1100}\}\\
\ket{\bm{\phi}_2}=\{\ket{0001},\ket{0101},\ket{1001},\ket{1101}\}\\
\ket{\bm{\phi}_3}=\{\ket{0010},\ket{0110},\ket{1010},\ket{1110}\}\\
\ket{\bm{\phi}_4}=\{\ket{0011},\ket{0111},\ket{1011},\ket{1111}\}
\eeql
It is easy to see, when any three of coefficient vectors $\bm{v}_i$ are equal to zero, the state must be two part $A_3\!A_4$ partially separable. Suppose that  non-zero coefficients belong to at least two different coefficient vectors, for example, 
\beq
\ket{\psi_{A_1\!A_2\!A_3\!A_4}}=\bm{v}_1\cdot\ket{\bm{\phi}_1}+\bm{v}_2\cdot\ket{\bm{\phi}_2}
\eeq
where at least two coefficients, which one belongs to $\bm{v}_1$ and another belongs to $\bm{v}_2$, are not zero. Without loss of generality, setting $x_1$ and $x_2$ are not zero, then since  
\beq
x_5=\frac{x_1x_6}{x_2},\quad x_9=\frac{x_1x_{10}}{x_2},\quad x_{13}=\frac{x_1x_{14}}{x_2}
\eeq
In other word, $\bm{v}_1$ is proportional to $\bm{v}_2$
\beq
\bm{v}_1=\frac{x_1}{x_2}\bm{v}_2
\eeq
Again note that
\beqa
\ket{\bm{\phi}_1}=\{\ket{00},\ket{01},\ket{10},\ket{11}\}\otimes\ket{00}\\
\ket{\bm{\phi}_2}=\{\ket{00},\ket{01},\ket{10},\ket{11}\}\otimes\ket{01}
\eeqa
Thus
\beqa
\ket{\psi_{A_1\!A_2\!A_3\!A_4}}&=&\frac{x_1}{x_2}\bm{v}_2\cdot\ket{\bm{\phi}_1}+\bm{v}_2\cdot\ket{\bm{\phi}_2}\nonumber\\
&=& \left(\bm{v}_2\cdot\{\ket{00},\ket{01},\ket{10},\ket{11}\}\right)\nonumber\\
& & \otimes\left(\frac{x_1}{x_2}\ket{00}+\ket{01}\right)
\eeqa
It implies that the state is two part $A_3A_4$ partially separable. In the similar way, we can prove the other cases and omit them in order to save space. While if $A_1A_3$ or $A_1A_4$ are two part partially separable, the proof can be carried out for the rearranged states. Therefore, our theorem has been proved to be valid for four partite systems with four qubits. 

For the systems more than four parts (four qubits), we have to use the mathematical induction to prove our theorem. However, all of statement is similar to the proof of one part partially separable case and so we do not intend writing down it in detail.    

In order to classify the states correctly, we also require that the separated two parts $A_i$ and $A_j$ are entangled, that is, $\bm{\xi}_{A_i}$ or $\bm{\xi}_{A_j}$ can not be equal to 1. Otherwise, both $A_i$ and $A_j$ should belong to one part partially separable classes. 

As an operational expression and for purpose of classification of states, we can write out our conclusion: for a pure state in $n$-partite systems with $n$ qubits (every part has one qubit), two part $A_i$ and $A_j$ (given $i$ and $j$ and setting $i<j$) partially separable sufficient and necessary condition is 
\beq
\label{2psc}
\bm{\xi}_{A_iA_j}^2=3/2
\eeq
where $\bm{\xi}_{A_i\!A_j}$ is a coherence vector of the reduced density matrix $\rho_{A_iA_j}=\Tr_{\prod_{k=1,k\neq i,j}^n A_k}(\rho_{A_1\!A_2\cdots A_n})$ (tracing off $n-2$ parts). Moreover, $\bm{\xi}_{A_i}^2$ or $\bm{\xi}_{A_j}^2$ is not equal to 1. 

For a $m$-part partially separable case, our proof method can be directly extended and used. Except for a little heavy computation, one does not needs more ideas and technology. W have enough physical reasons to conclude, in general

{\bf Corollary Three}  For a pure state in $n$-partite systems made up of $n$ qubits, $m$-partite $A_{j_1}A_{j_2}\cdots A_{j_m}$ subsystem partially separable necessary and sufficient condition is that the coherent vector of reduced density matrix of this $m$-partite subsystem (which includes the all of rearranged parts) has the maximum norm, that is
\beq
\label{pscgf}
\bm{\xi}_{A_{j_1}\!A_{j_2}\!\cdots\! A_{j_m}}^2=2\left(1-\frac{1}{2^m}\right)
\eeq
Of course, we also require that the separated $m$ parts are fully entangled in order to classify states clearly. 

Thus, we have finished our proof about our theorem. It has been seen that our proof only used some elementary and familiar ideas and methods. This is just one of our attempts, making this problem simpler and more intuitionistic, as well as easily operational. 

Obviously, the key conclusion of our theorem is the equivalence between the partial separability of the given $m$-partite subsystem and the purity of states of this $m$-partite subsystem. Our theorem and corollaries are very definite, clear and much simpler than the known ones. Moreover, they are easy to understand and operate. It must be emphasized that the state purity conditions of the given $m$-partite subsystem can be expressed by many equivalent forms. The ref.\cite{Miyake} may find a possible way among them, but his result seems to exist some unnecessary complication. Obviously, from his result, one does not obtain our theorem directly. In the operational aspect, our partial separability criterions have very obvious advantages. However, it is worthy to mentioning that the onion structure in ref.\cite{Miyake} is interesting and his/her method might contain more information one wants to know. We hope to study this problem in near future. 

Just because of simplicity and clearness of our theorem we can suggest an easily operational criterion of partial separability such as our corollaries. Our criterions can become a kinds of standard procedures to simplify greatly the task of classification of states.   

In fact, for a multipartite system, it is possible not enough only to classify the states into (fully) separable and entangled in order to understand and describe entanglement. It seems to be necessary, in our point of view, to classify the states to fully separable and non (fully) separable, and then non (fully) separable states are again divided into fully entangled, one part, two part, $\cdots$ partially separable classes because to quantify entanglement of fully entangled, different number of parts partially separable or partially entangled states might be different. 

Fully entangled states may be the most important resource for the applications including all of parts in quantum communications and quantum computing since all of ``cat states" belong to this class. If a $n$-partite system is $m$-part partially separable, we only need study respectively the entanglement in two divided $m$ and $n-m$ partite subsystems. Obviously, any partially separable states can not be transformed to the maximum entangled states by local quantum operation and classical communications (LOCC). In general, we have

{\bf Corollary Four} Under LOCC, it is invariant of partially separable property of the states in multipartite systems. 

From the definition of partially separable states and LOCC, it can be directly obtained. In other words, if a given state is $m$-part partially separable, then it keeps $m$-part partially separable property after carrying out any LOCC. We again see that separability, entanglement and distillation are closely interconnected together in quantum information.    

By using of our above idea and method, we can discuss the partially separable problems and carry out classification of states in the multipartite systems made up of qutrits, qudits as well as the multipartite systems made up of mixture of qubits, qutrits and qudits, which are arranged in our other papers (in prepare).

In the mixed states, the situation is a little complicated. If in a pure state decomposition $\rho=\sum_ip_i\rho^i$, we can find that each $\rho^i$ (pure state) is $m$-part partially separable, then this mixed state is $m$-part partially separable. However, in general, one can not find out all of kinds of pure state decompositions, how to finally and clearly determine the partially separable property of a given mixed state in a multipartite system is still difficult. Just as the fully separable problem of mixed states, the partially separable problem of mixed state in multipartite systems also opens now. 

The study is on progressing.

\end{document}